\documentclass[%showpacs, showkeys,
12pt,
preprint,preprintnumbers,nofootinbib,
groupedaddress,superscriptaddress,amsmath,amssymb]{revtex4}
\usepackage{graphicx}% Include figure files
\usepackage{dcolumn}% Align table columns on decimal point
\usepackage{bm}% bold math
\usepackage{amssymb}
\usepackage{amsmath}
\usepackage{epsfig}    
\usepackage{color}
\usepackage{slashed}
\usepackage{hhline}
\def\be{\begin{equation}}
\def\ee{\end{equation}}
\newcommand{\bea}{\begin{eqnarray}}
\newcommand{\eea}{\end{eqnarray}}
\newcommand{\nn}{\nonumber}

\numberwithin{equation}{section}

\begin{document}

{\begin{flushright}{KIAS-P17034}
\end{flushright}}

%%%%%%%%%
\title{Radiative neutrino mass \\
 in an alternative $U(1)_{B-L}$ gauge symmetry }
%\preprint{KIAS-P14078}
%

\author{Takaaki Nomura}
\email{nomura@kias.re.kr}
\affiliation{School of Physics, KIAS, Seoul 02455, Korea}

\author{Hiroshi Okada}
\email{macokada3hiroshi@cts.nthu.edu.tw}
\affiliation{Physics Division, National Center for Theoretical Sciences, Hsinchu, Taiwan 300}

\date{\today}

\begin{abstract}
We  propose a neutrino model in which neutrino masses are generated at one loop level and three right-handed fermions have non-trivial charges under $U(1)_{B-L}$ gauge symmetry in no conflict with anomaly cancellation. After the spontaneously symmetry breaking, a remnant $Z_2$ symmetry is induced and plays an role in assuring the stability of dark matter candidate.
\end{abstract}
\maketitle
\newpage

\section{Introduction}
%%%
Radiatively induced neutrino mass models are attractive candidate to explain the smallness of neutrino masses.
In such models, neutrino masses are not allowed at the tree level by some symmetries and they are generated at loop level.
Moreover dark matter (DM) candidate can easily be accommodated as a particle propagating inside a loop diagram generating the masses of neutrinos.
Based on these ideas, one loop induced neutrino models have widely been studied by a lot of authors; for example, see
refs.~\cite{a-zee, Cheng-Li, Pilaftsis:1991ug, Ma:2006km, Gu:2007ug, Sahu:2008aw, Gu:2008zf, AristizabalSierra:2006ri, Bouchand:2012dx, McDonald:2013hsa, Ma:2014cfa, Kajiyama:2013sza, Kanemura:2011vm, Kanemura:2011jj, Kanemura:2011mw, Schmidt:2012yg, Kanemura:2012rj, Farzan:2012sa, Kumericki:2012bf, Kumericki:2012bh, Ma:2012if, Gil:2012ya, Okada:2012np, Hehn:2012kz, Dev:2012sg, Kajiyama:2012xg, Toma:2013zsa, Kanemura:2013qva, Law:2013saa, Baek:2014qwa, Kanemura:2014rpa, Fraser:2014yha, Vicente:2014wga, Baek:2015mna, Merle:2015gea, Restrepo:2015ura, Merle:2015ica, Wang:2015saa, Ahn:2012cg, Ma:2012ez, Hernandez:2013dta, Ma:2014eka, Ma:2014yka, Ma:2015pma, Ma:2013mga, radlepton1, Okada:2014nsa, Brdar:2013iea, Okada:2015kkj, Bonnet:2012kz, Joaquim:2014gba, Davoudiasl:2014pya, Lindner:2014oea, Okada:2014nea, Mambrini:2015sia, Boucenna:2014zba, Ahriche:2016acx, Fraser:2015mhb, Fraser:2015zed, Adhikari:2015woo, Okada:2015vwh, Ibarra:2016dlb, Arbelaez:2016mhg, Ahriche:2016rgf, Lu:2016ucn, Kownacki:2016hpm, Ahriche:2016cio, Ahriche:2016ixu, Ma:2016nnn, Nomura:2016jnl, Hagedorn:2016dze, Antipin:2016awv, Nomura:2016emz, Gu:2016ghu, Guo:2016dzl, Hernandez:2015hrt, Megrelidze:2016fcs, Cheung:2016fjo, Seto:2016pks, Lu:2016dbc, Hessler:2016kwm, Okada:2015bxa,
Ko:2017quv, Ko:2017yrd, Lee:2017ekw, Antipin:2017wiz, Borah:2017dqx, Chiang:2017tai, Kitabayashi:2017sjz, Das:2017ski, Wang:2017mcy, Nomura:2017emk, Boehm:2006mi, He:2011hs, Farzan:2009ji, Herrero-Garcia:2017xdu, Suematsu:2009ww, Restrepo:2013aga}.  
%mainly focusses on the scenarios of one-loop level neutrino mass generation, and 
%%%%%%%%%%%%%%%%
In addition, refs. \cite{Cepedello:2017eqf, Wang:2016lve, Guo:2017ybk, Lindner:2016bgg, Cai:2017jrq} discuss the systematic analysis of (Dirac) neutrino oscillation, charged lepton flavor violation, and collider physics in the framework of neutrinophilic and inert two Higgs doublet model (THDM), respectively.

In many models, an additional discrete $Z_2$ symmetry has to be imposed in order to forbid tree level masses of neutrinos and to guarantee the stability of DM.
However $U(1)_{B-L}$ gauge symmetry can play such a role by taking non-trivial charge assignment of standard model (SM) gauge singlet fermions as shown in ref.~\cite{Singirala:2017see}, where 
%all the gauge anomalies can be canceled. Then particles
the lightest neutral particle with non-trivial $U(1)_{B-L}$ charge can be a DM candidate.
%%%
In this case, its stability is assured by a remnant $Z_2$ symmetry after the spontaneous $U(1)_{B-L}$ symmetry breaking.
%%%
Thus it is interesting to construct a radiative neutrino mass model based on the alternative charge assignment of $U(1)_{B-L}$. 

In this paper, we construct and analyze a model of $U(1)_{B-L}$ with alternative charge assignment, in which neutrino masses are generated at one loop level by introducing some exotic scalar fields.
%%%%%%%
Also we consider a physical Goldstone boson (GB), which is induced as a consequence of global symmetry in our scalar potential 
 introducing two types of SM gauge singlet scalar fields with nonzero $U(1)_{B-L}$ charges and vacuum expectation values (VEVs). 
%global symmetry in the scalar potential associated with $\varphi_{1,2}$.  Note that we have freedom to identify  which component of $(z'_{\varphi_1},z'_{\varphi_2})$ is the GB, and we choose $G\equiv z'_{\varphi_1}$ to be GB in our analysis.
We provide formulas of neutrino mass matrix, decay ratio of lepton flavor violating process and relic density of our DM candidate that is determined by interactions associated with the physical GB and an additional vector gauge boson $Z'$ from $U(1)_{B-L}$.
Then numerical global and benchmark analyses are carried out to search for parameter sets that can fit the neutrino oscillation data and satisfy experimental constraints of lepton flavor violations (LFVs) and relic density of DM.
%%%%%%

This paper is organized as follows.
In Sec.~II, we show our model, %to introduce exotic fermions and bosons with some additional symmetries,  
and formulate the neutral fermion sector, boson sector, lepton sector, and dark matter sector.
Also we analyze the relic density of DM without conflict of direct detection searches, and carry out global analysis.
Finally We conclude and discuss in Sec.~III.
%\newpage

%%%%%%%%%%%%%%%%%%%%%%%%%%%%%%%%%%%%%
%\section{The Model}
%\subsection{Model setup}

 \begin{widetext}
\begin{center} 
\begin{table}[t]%[tbc]
%\begin{tiny}
\begin{tabular}{|c||c|c|c|c|c||c|c|c|}\hline\hline  
%&\multicolumn{5}{c||}{SM leptons} & \multicolumn{3}{c|}{Exotic fermions} \\\hline
Fermions& ~$Q_L$~ & ~$u_R$~ & ~$d_R$~ &~$L_L$~ & ~$e_R$~ & ~$N_{R_1}$~ & ~$N_{R_2}$~ & ~$N_{R_3}$~ 
\\\hline 
$SU(3)_C$ & $\bm{3}$  & $\bm{3}$  & $\bm{3}$  & $\bm{1}$  & $\bm{1}$  & $\bm{1}$  & $\bm{1}$  & $\bm{1}$  \\\hline 
 %%%
 $SU(2)_L$ & $\bm{2}$  & $\bm{1}$  & $\bm{1}$ & $\bm{2}$ & $\bm{1}$  & $\bm{1}$ & $\bm{1}$ & $\bm{1}$   \\\hline 
 %%%
$U(1)_Y$ & $\frac16$ & $\frac23$  & $-\frac{1}{3}$ & $-\frac12$  & $-1$ & $0$ & $0$ & $0$    \\\hline
 %%%
 $U(1)_{B-L}$ & $\frac13$ & $\frac13$  & $\frac13$ & $-1$  & $-1$   & $-4$   & $-4$   & $5$   \\\hline
 %%%
\end{tabular}
%%%%%%%%%%%
\begin{tabular}{|c||c|c|c|c|c|}\hline\hline
  Bosons  &~ $H$  &~ $\eta$  ~ &~ $s$~ &~ $\varphi_1$ &~ $\varphi_2$ \\\hline
$SU(3)_C$ & $\bm{1}$  & $\bm{1}$  & $\bm{1}$  & $\bm{1}$ & $\bm{1}$ \\\hline 
$SU(2)_L$ & $\bm{2}$ & $\bm{2}$  & $\bm{1}$ & $\bm{1}$ & $\bm{1}$  \\\hline 
$U(1)_Y$ & $\frac12$ & $\frac12$  & $0$ & $0$ & $0$    \\\hline
 $U(1)_{B-L}$ & $0$ & $-3$ & $4$ & $1$  & $8$  \\\hline
\end{tabular}%
%%%%%%%%%%%
\caption{Field contents of fermions and bosons
and their charge assignments under $SU(3)_C\times SU(2)_L\times U(1)_Y\times U(1)_{B-L}$, where flavor indices are abbreviated.}
\label{tab:1}
% \end{tiny}
\end{table}
\end{center}
\end{widetext}

\if0
\begin{table}[t]
\centering {\fontsize{10}{12}
\begin{tabular}{|c||c|c|c|c|c|}\hline\hline
  Bosons  &~ $H$  &~ $\eta$  ~ &~ $s$~ &~ $\varphi_1$ &~ $\varphi_2$ \\\hline
$SU(3)_C$ & $\bm{1}$  & $\bm{1}$  & $\bm{1}$  & $\bm{1}$ & $\bm{1}$ \\\hline 
$SU(2)_L$ & $\bm{2}$ & $\bm{2}$  & $\bm{1}$ & $\bm{1}$ & $\bm{1}$  \\\hline 
$U(1)_Y$ & $\frac12$ & $\frac12$  & $0$ & $0$ & $0$    \\\hline
 $U(1)_{B-L}$ & $0$ & $-3$ & $4$ & $1$  & $8$  \\\hline
\end{tabular}%
} 
\caption{Boson sector. }
\label{tab:2}
\end{table}
\fi

\section{ Model setup and phenomenologies}
In this section, we introduce our model.
First of all, we impose an additional $U(1)_{B-L}$ gauge symmetry with  three right-handed neutral fermions $N_{R_i}(i=1-3)$,
where the right-handed neutrinos have $U(1)_{B-L}$ charge $-4$, $-4$ and $5$. Then all the anomalies we have to consider are  $U(1)_{B-L}^3$, and $U(1)_{B-L}$, which are found to be zero~\cite{Singirala:2017see}.
On the other hand, even when we introduce two types of isospin singlet bosons $\varphi_1$ and $\varphi_2$ in order to acquire nonzero Majorana masses after the spontaneous symmetry breaking of  $U(1)_{B-L}$, one cannot find active neutrino masses due to the absence of Yukawa term $\bar L_L \tilde H N_R$.
Thus we introduce an isospin singlet and doublet inert bosons $s$ and $\eta$ with nonzero $U(1)_{B-L}$ charges, and neutrino masses are induced at one-loop level as shown in Fig.~\ref{fig:diagram}. Also the stability of DM  is assured by a remnant $Z_2$ symmetry at renormalizable level after the spontaneous breaking  where $N_{R_i}$, $\eta$ and $s$ are $Z_2$ odd and the other fields are $Z_2$ even~\footnote{At non-renormalizable level we would have $Z_2$ breaking term inducing decay of DM such as $\bar L N_{R_{1,2}} H \varphi_1^3$. Such a term is suppressed by cut-off scale and we can assume DM is sufficiently long-lived.}.
Field contents and their assignments for fermions and bosons are respectively given by Table~\ref{tab:1}.
% Notice here that all the anomalies related to $U(1)_Y$; $U(1)_Y^2\times U(1)_{B-L}$ and  $U(1)_Y\times U(1)_{B-L}^2$, are zero. 
%\subsection{Yukawa interactions and scalar sector}{\it Yukawa Lagrangian}:
Under these symmetries, the renormalizable Lagrangian for lepton sector and Higgs potential are respectively given by 
\begin{align}
-{\cal L}_{L}&=
%\nn\\&
 (y_\ell)_{a b}\bar L_{L_a} e_{R_b} H + (y_\nu)_{ai}\bar L_{L_a} \tilde\eta N_{R_i}
+y_{N_{i3}} \bar N^C_{R_i}  N_{R_3} \varphi_1^*+ y'_{N_{ij}} \bar N^C_{R_i}  N_{R_j} \varphi_2 
+{\rm c.c.},\\
%%%
V&= \mu_H^2 H^\dag H + \mu_\eta^2 \eta^\dag \eta+ \mu_s^2 s^*s + \mu^2_{\varphi_1}\varphi_1^*\varphi_1  + \mu^2_{\varphi_2}\varphi_2^*\varphi_2\nn\\
&+\mu (s^2\varphi_2^*+{\rm c.c.})+\lambda_0(H^\dag\eta s\varphi_1^* + {\rm c.c.})\nn\\
%%%
&+
\lambda_H (H^\dag H)^2+\lambda_\eta (\eta^\dag \eta)^2 +\lambda_s (s^* s)^2 + \lambda_{\varphi_1}(\varphi_1^*\varphi_1)^2 + \lambda_{\varphi_2}(\varphi_2^*\varphi_2)^2+\lambda_{H\eta} (H^\dag H)(\eta^\dag\eta)\nn\\
&+
\lambda'_{H\eta} (H^\dag \eta)(\eta^\dag H) + \lambda_{Hs} (H^\dag H)(s^*s)
+
 \lambda_{H\varphi_1} (H^\dag H)(\varphi_1^*\varphi_1)+ \lambda_{H\varphi_2} (H^\dag H)(\varphi_2^*\varphi_2)
\nn\\
&+
\lambda_{\eta s} (\eta^\dag \eta)(s^*s)+ \lambda_{\eta\varphi_1} (\eta^\dag \eta)(\varphi_1^*\varphi_1)+ \lambda_{\eta\varphi_2} (\eta^\dag \eta)(\varphi_2^*\varphi_2)
+ \lambda_{s\varphi_1} (s^*s)(\varphi_1^*\varphi_1)\nn\\
&+
\lambda_{s\varphi_2} (s^*s)(\varphi_2^*\varphi_2)
+ \lambda_{\varphi_1\varphi_2} (\varphi_1^*\varphi_1) (\varphi_2^*\varphi_2)
\label{eq:lag-lep}
\end{align}
where $\tilde H \equiv (i \sigma_2) H^*$ with $\sigma_2$ being the second Pauli matrix, $(a,b)$ runs over $1$ to $3$, and $(i,j)$ runs over $1$ to $2$.
%%%%%%%%%
\subsection{ Scalar sector}
The scalar fields are parameterized as 
\begin{align}
%\begin{tiny}
&H =\left[\begin{array}{c}
w^+\\
\frac{v + h +i z}{\sqrt2}
\end{array}\right],\quad 
%%%
\eta =\left[\begin{array}{c}
\eta^+\\
\frac{ \eta_R +i \eta_I}{\sqrt2}
\end{array}\right],\quad 
%%%
s=
\frac{s_R + is_I }{\sqrt2},\quad 
%%%
\varphi_i=
\frac{v'_i+\varphi_{R_i} + iz'_{\varphi_i}}{\sqrt2},\ (i=1,2),
\label{component}
%\end{tiny}
\end{align}
where $w^+$ and $z$ are absorbed by the SM gauge bosons $W^+$ and $Z$ as Nambu-Goldstone boson (NGB), 
and one of the massless CP odd boson after diagonalizing the matrix in basis of $(z'_{\varphi_1},z'_{\varphi_2})$ with nonzero VEVs is  absorbed by the $B-L$ gauge boson $Z'$. \\
{\it CP-odd scalar} ($Z_2$ even) :
 As a result, one physical massless CP-odd GB is induced, which is due to a breaking of global symmetry in the scalar potential associated with $\varphi_{1,2}$; an golobal $U(1)$ symmetry under which $\varphi_1$ and $\varphi_2$ transforms separately. 
Note that we have freedom to identify  which component of $(z'_{\varphi_1},z'_{\varphi_2})$ is the GB, and
we choose $G\equiv z'_{\varphi_1}$ to be GB in our analysis. 
%%%%
One can identify the CP-odd boson of $\varphi_1$ as NGB, when $v'_{2}<<v'_{1}$.
Here we consider  the CP-odd boson of $\varphi_2$ is a physical GB; $z'_{\varphi_2}$, and
it contributes to phenomenologies such as DM.
We also note that the existence of this physical Goldstone boson does not cause serious problem in particle physics or cosmology since it does not interact with SM particles directly and decouples from thermal bath in early Universe.
%%%%
Also we assume that coupling between $G$ and SM Higgs is negligibly small by choosing parameters in scalar potential, and GB does not affect phenomenology; 
the contribution to relativistic degrees of freedom by GB is also small since it decouples in early stage of the universe due to small interactions.
%The remaining nonzero mass and its mixing matrix  can be given by
%\begin{align}m_a^2=...,\quadO_I=...\end{align}
%%%
{\it CP-even scalar}:
Inserting tadpole conditions, the CP even matrix in basis of $(\varphi_{R_1},\varphi_{R_2}, h)$ with nonzero VEVs is given by 
\begin{align}
M_R^2
&\equiv
\left[\begin{array}{ccc}
2 v'^2_1\lambda_{\varphi_1} &  v'_1 v'_2\lambda_{\varphi_1\varphi_2} &  v v'_1\lambda_{H\varphi_1} \\ 
v'_1 v'_2\lambda_{\varphi_1\varphi_2} & 2 v'^2_2 \lambda_{\varphi_2} &  v v'_2\lambda_{H\varphi_2} \\ 
 v v'_1\lambda_{H\varphi_1} & v v'_2\lambda_{H\varphi_2} & 2 v^2 \lambda_{H} \\ 
\end{array}\right],
\end{align}
where we define the mass eigenstate $h_{i}$ ($i=1-3$), and mixing matrix $O_R$ to be $m_{h_{i}}=O_R M_R^2 O_R^T$ and $(\varphi_{R_1},\varphi_{R_2},  h)^T=O_R^T h_i$. Here $h_3\equiv h_{SM}$ is the SM Higgs, therefore,  $m_{h_3}=$125 GeV.
 In addition, we assume mixing among SM Higgs and other CP-even scalars are small to avoid experimental constraints for simplicity. 
  %%%%%%%%%%%%%%%%%%%
  
{\it The inert scalar sector}: we obtain mass matrix in the basis of $( s_{R(I)}, \eta_{R(I)})$ such as  
\begin{equation}
M^2_{s_{R(I)} \eta_{R(I)}} = \frac12 \begin{pmatrix}
v'^2_1 \lambda_{s \varphi_1} +v^2\lambda_{Hs}+\lambda_{s \varphi_2} v'^2_2+2\mu_s^2  & (-)\lambda_0 v'_1 v \\
(-)\lambda_0 v'_1 v  & v'^2_1 \lambda_{\eta \varphi_1} +v^2(\lambda_{H\eta}+\lambda'_{H\eta})+ v'_2\lambda_{\eta \varphi_2}+2\mu_\eta^2
\end{pmatrix}.
\end{equation}
In our analysis, we assume $\lambda_0 \ll 1$ so that mixing between $s_{R(I)}$ and $\eta_{R(I)}$ is small, and we apply mass insertion approximation in calculating neutrino mass matrix below. 
Thus each of mass eigenvalues as a leading order is given by
\begin{align}
m_{s_{R(I)}}^2 \approx m_{s}^2&\equiv  \frac{v'^2_1 \lambda_{s \varphi_1} +v^2\lambda_{Hs}+\lambda_{s \varphi_2} v'^2_2+2\mu_s^2 }{2},\\
%%%
m_{\eta_{R(I)}} \approx m_{\eta}^2&\equiv  \frac{v'^2_1 \lambda_{\eta \varphi_1} +v^2(\lambda_{H\eta}+\lambda'_{H\eta})+ v'_2\lambda_{\eta \varphi_2}+2\mu_\eta^2 }{2},
  \end{align}
 where we omit the mixing effect as an approximation. 
 For charged scalar $\eta^\pm$, we have no mixing effect and its mass is simply given by 
 \begin{equation}
 m_{\eta^\pm} = m_{\eta}.
 \end{equation}
 Thus we have degenerated mass eigenvalues for components of inert doublet $\eta$ in our approximation.

 {\it Stability of the potential:} The global minimum at $\langle\eta\rangle=\langle s\rangle=0$ requires the following conditions~\cite{Belanger:2012vp}: 
  \begin{align}
 & 0<(\lambda_H,\ \lambda_\eta,\ \lambda_s,\ \lambda_{\varphi_1},\  \lambda_{\varphi_2},\ \lambda_{\eta s},\ \lambda_{Hs},\ \lambda_{s\varphi_1},\ \lambda_{s\varphi_2},\ \lambda_{H\eta}+\lambda'_{H\eta},\ \lambda_{\eta\varphi_1}+\lambda'_{\eta\varphi_2}),
 \\&
 0<\mu v_{\varphi_2},\  
 0<\sqrt{\lambda_{Hs} \lambda_{\eta\varphi_1}}+\frac{\lambda_0}{3},\ 
  0<\sqrt{(\lambda_{H\eta}+\lambda'_{H\eta}) \lambda_{s\varphi_1}}+\frac{\lambda_0}{3},\ 
  0<\sqrt{\lambda_{H\varphi_1} \lambda_{\eta s}}+\frac{\lambda_0}{3}.
  \end{align}

 {\it Physical Goldstone boson}: Here we also discuss decoupling of the physical GB from thermal bath where we assume it is thermalized via Higgs portal interaction following discussion in ref.~\cite{Weinberg:2013kea}. 
Note that $Z'$ interaction is subdominant since $Z'$ is heavy and gauge coupling should be small from collider constraints as we discuss later.
 The effective interaction among our GB $z'_{\varphi_2}$ and the SM fermions is induced from the interactions $-1/(2v'_2) \varphi_{R_2} \partial_\mu z'_{\varphi_2} \partial^\mu z'_{\varphi_2}$, $\lambda_{H \varphi_2} v'_2 v \varphi_{R_2} h$ and the SM Yukawa interactions such as: 
 \begin{equation}
 - \frac{\lambda_{H \varphi_2} m_f}{2 m_{\varphi_{R_2}}^2 m_{h}^2} \partial_\mu z'_{\varphi_2} \partial^\mu z'_{\varphi_2} \bar f f,
 \end{equation}
 where $m_f$ is the mass of the SM fermion $f$, $m_h$ is the SM Higgs mass, and we take $\varphi_{R_2}$ as mass eigenstate for simplicity.
 The temperature at which $z'_{\varphi_2}$ decouples from thermal bath is roughly calculated by~\cite{Weinberg:2013kea} 
 \begin{equation}
 \frac{\text{collision rate}}{\text{expansion rate}} \simeq \frac{\lambda_{H \varphi_2}^2 m_f^2 (k T)^5 m_{PL}}{m_{\varphi_{R_2}}^4 m_{h}^4} \sim 1,
 \end{equation}
 where $m_{PL}$ denotes the Planck mass and $m_f$ should be smaller than $kT$ so that $f$ is in thermal bath. The decoupling temperature is then estimated as
 \begin{equation}
 k T \sim 4.8 \, {\rm GeV} \left( \frac{m_{\varphi_{R_2}}}{100 \, {\rm GeV}} \right)^{\frac{4}{5}} \left( \frac{{\rm GeV}}{m_f} \right)^{\frac{2}{5}} \left( \frac{0.01}{\lambda_{H \varphi_2}} \right)^{\frac{2}{5}}.
 \end{equation}
 Thus $z'_{\varphi_2}$ can decouple from thermal bath sufficiently early and does not contribute to the effective number of active neutrinos~\cite{Brust:2013xpv}.

\subsection{Gauge sector} 
After $U(1)_{B-L}$ symmetry breaking we have massive $Z'$ boson.
In this model, $Z'$--$Z$ mixing could be induced only through kinetic mixing effect since Higgs doublet does not have $B-L$ charge.
Here we assume the kinetic mixing is negligibly small and we can avoid constraint from mixing effect. 
The mass of $Z'$ is then given by 
\begin{equation}
m_{Z'} = g_{BL} \sqrt{v'^2_1 + 64 v'^2_2} \simeq g_{BL} v'_1,
\end{equation}
where we have applied $v'_1 \gg v'_2$ for approximation.
As we see below collider constraint indicates $U(1)_{B-L}$ breaking scale is $m_{Z'}/g_{BL} > 10$ TeV.

\subsection{Fermion Sector}
%%%
The mass matrix for the neutral fermions in basis of $N_{R_{1,2,3}}$, and given by
\begin{align}
M_N=\frac{1}{\sqrt2}
\left[\begin{array}{ccc}
y'_{N_{11}} v'_2 & y'_{N_{12}} v'_2  & y_{N_{13}} v'_1   \\
y'_{N_{12}} v'_2 & y'_{N_{22}} v'_2  & y_{N_{23}} v'_1   \\
y_{N_{13}} v'_1 & y_{N_{23}} v'_1  & 0   \\
\end{array}\right],
\label{eq-Nmass}
\end{align}
where we define this matrix is diagonalized by 3 by 3 orthogonal  matrix $V_N$ as $ M_{\psi_i}  \equiv (V_N M_N V_N^T)_i$ $i=1\sim3$, where  $M_{\psi_i}$ is the mass eigenvalue.
The mass eigenstates are given by $\psi_{i} = (V_N)_{ij} N_{R_j}$.

%%%%%%%%%%%%%%%%%%%
\begin{figure}[t]
\begin{center}
\includegraphics[width=70mm]{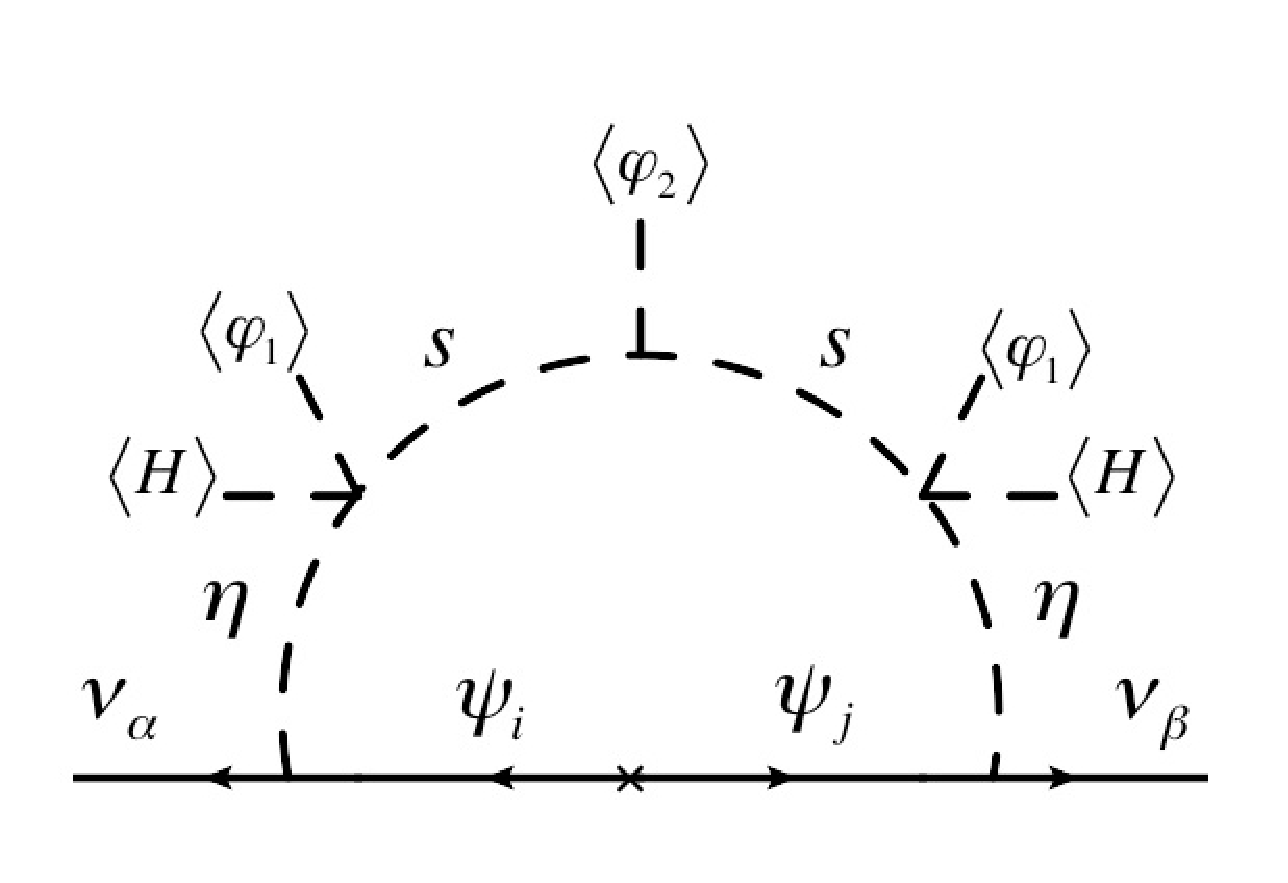} 
\caption{The one loop diagram which induces neutrino masses. } 
  \label{fig:diagram}
\end{center}\end{figure}
%%%%%%%%%%%%%%%%%%%

  \subsection{Lepton sector and lepton flavor violations}
The charged lepton masses are given by $m_\ell =y_\ell v/\sqrt2$ after the electroweak symmetry breaking, where $m_\ell$ is assumed to be the mass eigenstate.
 %%%%%%%%%
{The neutrino mass matrix is induced at the one-loop level as shown in Fig.~\ref{fig:diagram}, and its mass-insertion-approximation form is given by}
\footnote{Notice here that our one-loop function is different from the one of Ma model~\cite{Ma:2006km}, since we apply a mass insertion approximation method.}
  \begin{align}
&({\cal M}_{\nu})_{\alpha\beta}  = \frac{(\lambda_0 v v'_1)^2 \mu v'_2}{4\sqrt2 (4\pi)^2}
\frac{{(Y_\nu)_{\alpha i} M_{\psi_i} (Y_\nu^T)_{i\beta} }}{m_s^6} F_\nu(r_{\eta},r_{\psi_i}),\nn \\
%%%
& F_\nu(r_{1},r_{2}) =\nn\\
& \frac{  (1-r_1)(1+r_1-2r_2)(r_1-r_2)(1-r_2)-(1-r_2)^2[r_2+r_1(-2r_1+r_2)]\ln[r_1] + (1-r_1)^3r_2\ln[r_2]}
 {2(1-r_1)^3(1-r_2)^2 (r_1-r_2)^2},
%\int_0^1 dx \int_0^{1-x} dy \frac{xy}{[x m_s^2 + y m_{\eta_0}^2 +(1-x-y) M_{\psi_i}^2]^3},
  \end{align}
  where $r_i\equiv (m_i/m_s)^2$, $(Y_\nu)_{\alpha i}\equiv \sum_{j=1}^3\frac{(y_\nu)_{\alpha j} (V_N^T)_{ji}}{\sqrt2}$.
 Once we define $D_\nu\equiv U_{MNS} {\cal M}_{\nu} U_{MNS}^T \equiv U_{MNS} (Y_\nu R Y_\nu^T)  U_{MNS}^T$,
 $Y_\nu$ can be rewritten in terms of observables and several arbitral parameters as:
 \begin{align}
 Y_\nu = U^\dag_{MNS} D_\nu^{1/2} O R^{-1/2},\quad 
 R_{ii} \equiv \frac{(\lambda_0 v v'_1)^2 \mu v'_2 M_{\psi_i}}{4\sqrt2 (4\pi)^2 m_s^6} F_\nu(r_{\eta},r_{\psi_i}),
  \end{align}
where $O\equiv O(\theta_1,\theta_2,\theta_3)$, satisfying $OO^T=1$, is an arbitral 3 by 3 orthogonal matrix with complex values, and $U_{MNS}$ and $D_\nu$ are measured in~\cite{Gonzalez-Garcia:2014bfa}.
 Here typical order of $R_{ii}$ is shown as 
\begin{equation}
R_{ii} \sim 3.0 \times 10^{-2} \left( \frac{\rm TeV}{m_s} \right)^6 \left( \frac{v'_1}{\rm 10 \, TeV} \right)^2 \left( \frac{\mu}{\rm TeV} \right) 
\left( \frac{v'_2}{\rm TeV} \right) \left( \frac{M_{\psi_i}}{\rm TeV} \right)  \lambda_0^2 \ {\rm GeV}, 
\end{equation}
where we have taken loop factor $F_\nu$ to be $\mathcal{O}(1)$ for simplicity.
Taking $\lambda_0 = 0.01(0.1)$, we obtain $Y_\nu \lesssim 10^{-2(4)}$ since order of neutrino mass is $\mathcal{O}(10^{-10})$ GeV.

{\it Lepton flavor violations}: LFV processes $\ell \to \ell' \gamma$ are induced from the neutrino Yukawa couplings at one-loop level, and their forms are given by
\begin{align}
BR(\ell_\alpha\to \ell_\beta \gamma)&\approx\frac{4\pi^3\alpha_{em}C_{\alpha\beta}}{3(4\pi)^4G_F^2}
\left|\sum_{i=1}^3(Y_\nu^\dag)_{\beta i} (Y_\nu)_{i\alpha} F_{lfv}(\psi_i,\eta^\pm)\right|^2,\\
%%%
F_{lfv}(a,b)&\equiv\frac{2 m_a^6+3m_a^4m_b^2-6m_a^2m_b^4+m_b^6+12m_a^4m_b^2\ln\left[\frac{m_b}{m_a}\right]}{(m_a^2-m_b^2)^4},
\end{align}
where $\alpha_{em}\approx1/137$ is the fine-structure constant, $G_F\approx1.17\times10^{-5}$ GeV$^{-2}$ is the Fermi constant,
and $C_{21}\approx1$, $C_{31}\approx 0.1784$, $C_{32}\approx0.1736$. 
Experimental upper bounds are found to be~\cite{TheMEG:2016wtm, Adam:2013mnn}: 
\begin{equation}
{\rm BR}(\mu\to e \gamma)\lesssim 4.2\times 10^{-13},\ 
{\rm BR}(\tau\to e \gamma)\lesssim 3.3\times 10^{-8},\ 
{\rm BR}(\tau\to \mu \gamma)\lesssim 4.4\times 10^{-13},
 \end{equation}
where we define $\ell_1\equiv e$,  $\ell_2\equiv \mu$, and  $\ell_3\equiv \tau$. Notice here that muon $g-2$ is negatively induced
that conflicts with the current experimental data. 
%Now we show a benchmark set to satisfy neutrino oscillation data and LFVs below.
%------------------------------------------------------------------------
\begin{figure}[t]
\centering
\includegraphics[width=7cm]{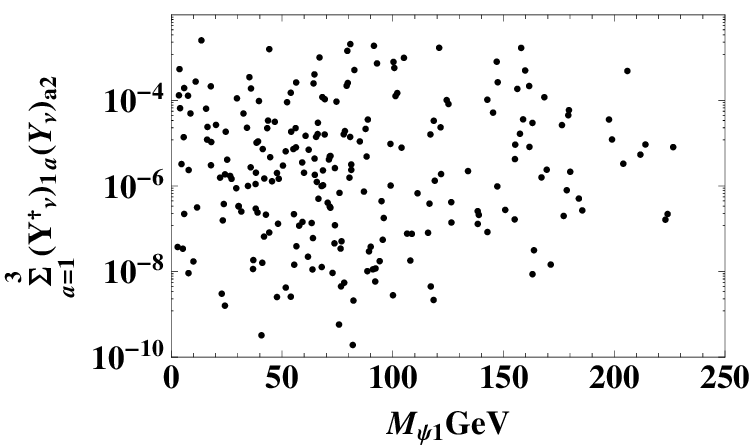}
\includegraphics[width=7cm]{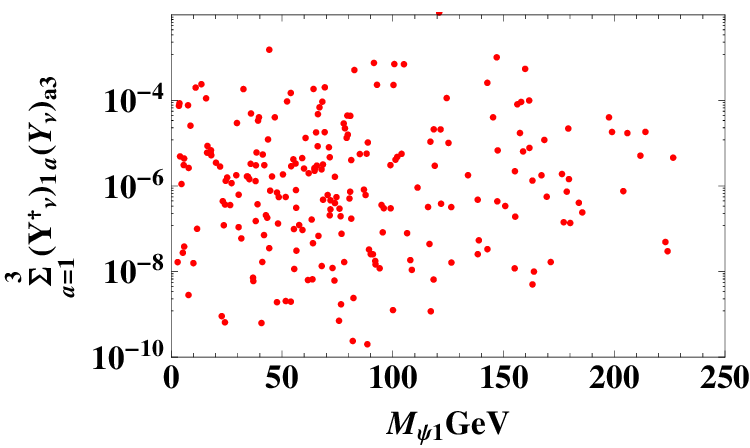}
\includegraphics[width=7cm]{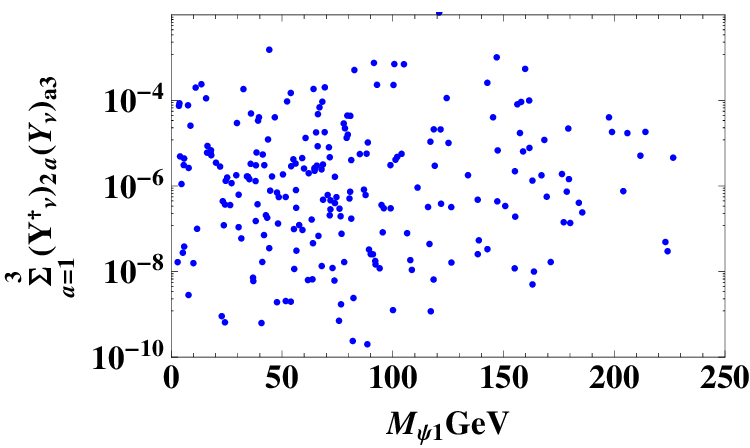}
\caption{The correlation between Yukawa coupling squared related to the LFV and the lightest neutral fermion mass $M_{\psi_1}$.  }
\label{fig:neut-lfvs}
\end{figure}
%------------------------------------------------------------------------

\begin{figure}[t]
\centering
\includegraphics[width=7cm]{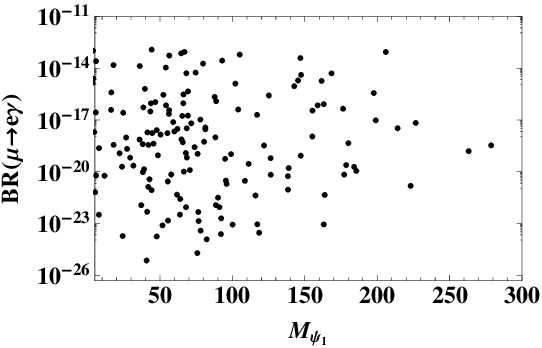}
\includegraphics[width=7cm]{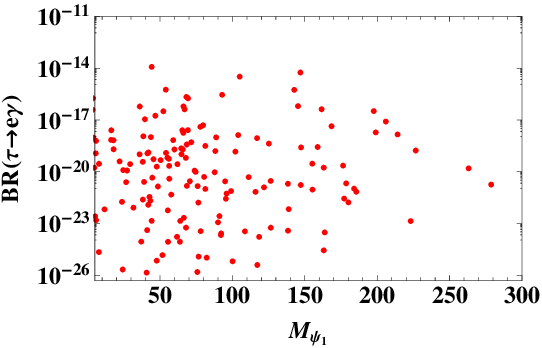}
\includegraphics[width=7cm]{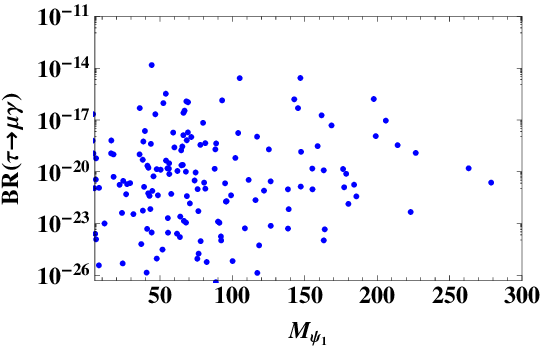}
\caption{The correlation between the LFV branching ratios and the lightest neutral fermion mass $M_{\psi_1}$.  }
\label{fig:lfvs}
\end{figure}

 Here we scan some parameters and derive allowed parameter region. The parameter ranges are chosen as
\begin{equation}
m_\eta \in [100, 1000] \ {\rm GeV}, \quad (M_N)_{ij} \in [100, 10000] \ {\rm GeV}, \quad m_s \in [100, 1000] \ {\rm GeV},
\end{equation}
where we fix $\mu = 100$ GeV and $\lambda_0 v'_1 = 575$ GeV.
We then search for Yukawa couplings $(Y_\nu)_{ij}$ which can accommodate with neutrino oscillation data and satisfy LFV constraints. 
Note also that we take degenerate mass for neutral and charged component of $\eta$ to avoid constraints from oblique parameters. 
In figs.~\ref{fig:neut-lfvs}, we show the global analysis to satisfy the neutrino oscillation data and LFVs in terms of the lightest neutral fermion mass $M_{\psi_1}$, (which is identified as a DM candidate in the next subsection), and each of Yukawa coupling squared related to the LFV,
where all the input region parameters include what we use the analysis of DM below.
The black points show the allowed region from ${\rm BR}(\mu\to e \gamma)$, the red points show the one from ${\rm BR}(\tau\to e \gamma)$, 
and , the blue points show the one from ${\rm BR}(\tau\to \mu \gamma)$.
All these constraints suggest that each of Yukawa coupling squared are of the order $10^{-4}$ at most.
 In Fig.~\ref{fig:lfvs}, we also show LFV BRs as functions of $M_{\psi_1}$. 
We find that $BR(\mu \to e \gamma)$ tends to be slightly larger than the other BRs while $BR(\tau \to e \gamma)$ and $BR(\tau \to \mu \gamma)$ have almost the same behavior. 
%%%%%%%%%%%%%%%%%%%%%%

%------------------------------------------------------------------------
%------------------------------------------------------------------------

\subsection{$Z'$ boson production at the LHC}
Here we discuss collider physics of $Z'$ boson in the model.
Our $Z'$ can be produced at the LHC since it couples to quarks due to $B-L$ charge.
Basically $Z'$ can decay into particles with $U(1)_{B-L}$ charge if kinematically allowed.
In our scenario, we consider decay modes of SM fermion pair, DMs, and $\varphi_{R_2} z'_{\varphi_2}$ where we consider negligibly small mixing in CP-even scalar sector and we assume the other modes are kinematically forbidden.
Notice that, for DM, we focus on the lightest inert fermion $\psi_1$, defining $\psi_1 \equiv X$ and $M_{\psi_1}\equiv M_X$.
The gauge interactions of $Z'$ are given by
\begin{align} 
\mathcal{L}_{Z'} =& \frac{g_{BL}Q^X_{BL}}{2} \bar X\gamma^\mu \gamma^5 X Z'_\mu +g_{BL} Q^f_{BL} \bar f_{SM} \gamma^\mu f_{SM} Z'_\mu \nn \\
& - g_{BL} \bar \nu\gamma P_L \nu Z'_\mu +i 8g_{BL} Z'^\mu (\partial_\mu \varphi_{R_2} z'_{\varphi_2} -  \varphi_{R_2} \partial_\mu z'_{\varphi_2} ),
\end{align} 
where $g_{BL}$ is $B-L$ gauge coupling, $Q^X_{BL}\equiv-4+9 (V_N^*)_{13} (V_N^T)_{31}$ applying unitary condition $V_N^\dag V_N=1$ ,  $Q^f_{BL}$ is the charge of $B-L$ symmetry for SM fermion $f_{SM}$.
Then the partial decay widths are obtained as
\begin{align}
& \Gamma_{Z' \to \bar f_{SM} f_{SM}} = X_f \frac{(Q_{BL}^fg_{BL})^2}{12 \pi} \left( 1 + \frac{2 m_{f_{SM}^2}}{m_{Z'}^2} \right) \sqrt{1 - \frac{4 m_{f_{SM}}^2}{m_{Z'}^2}}, \label{eq:width1} \\
& \Gamma_{Z' \to  X X} = \frac{(Q_{BL}^X g_{BL})^2}{24 \pi} \left( 1 + \frac{2 M_{X}^2}{m_{Z'}^2} \right) \sqrt{1 - \frac{4 M_{X}^2}{m_{Z'}^2} }, \label{eq:width2} \\
& \Gamma_{Z' \to \varphi_{R_2} z'_{\varphi_2}} = \frac{4 g_{BL}^2}{3 \pi} m_{Z'} \left( 1 - \frac{m_{\varphi_{R_2}}^2 }{m_{Z'}^2} \right)^3 \label{eq:width3},
\end{align}
where $X_f = 1/2(1)$ for SM neutrinos (charged leptons and quarks).

\begin{figure}[t]
\begin{center}
\includegraphics[width=70mm]{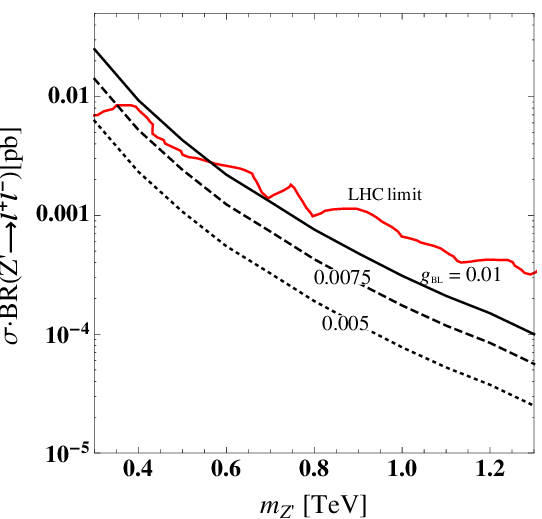} \qquad
\caption{The product of $Z'_R$ production cross section and $BR(Z'_R \to \ell^+ \ell^-)$ where region above red curve is excluded by the latest data~\cite{Aaboud:2017buh}. } 
  \label{fig:ZpLHC}
\end{center}\end{figure}

We estimate the $Z'$ production cross section using {\it CalcHEP}~\cite{Belyaev:2012qa} by use of the CTEQ6 parton distribution functions (PDFs)~\cite{Nadolsky:2008zw}, implementing relevant interactions.
In Fig.~\ref{fig:ZpLHC}, we show $\sigma(pp \to Z') BR(Z' \to \ell^+ \ell^-)$ with $\ell = \mu, e$ as a function of $m_{Z'}$ applying $m_{\varphi_{R_2}} = 200$ GeV and $m_X = 250$ GeV, 
which is compared with the current LHC limit.
We find that gauge coupling $g_{BL}$ should be $\mathcal{O}(0.01)$ or smaller to avoid the constraint when $m_{Z'} \lesssim 1$ TeV.

%%%%%%%%%%%%%%%%%%%
%%%%%%%%%%%%%%%%%%

%%%%%%%%%%%%%%%%%%%%%%

\subsection{ Dark matter} 
 In our scenario, we will focus on the lightest inert fermion $\psi_1$ as the DM candidate, $\psi_1 \equiv X$ and $M_{\psi_1}\equiv M_X$, as we discussed in previous subsection.
Note also that we can have bosonic DM candidate although we omit the discussion in this paper~\footnote{The bosonic DM candidate has been discussed in ref.\cite{Singirala:2017see}.}.
{\it Firstly, we assume contribution from the Higgs mediating interaction is negligibly small and DM annihilation processes are dominated by the gauge interaction with $Z'$; we thus can easily avoid the constraints from direct detection searches such as LUX~\cite{Akerib:2016vxi}.} 
%%%
{One might think nonzero contributions to the direct detection from the interaction via $Z'$ which would give strong constraint on the gauge coupling.
However constraint from DM direct detection is not significant in our model since the vector current of DM, 
which induces spin independent cross section, identically vanishes due to the Majorana property of our DM $X$.
Contribution to the direct detection arises from only vector axial current (as we will show below), which does not give the nonzero spin independent cross section but
spin dependent one. Therefore we do not need to consider the direct detection constraints, since all of them are safe.\\
%%%
{\it Relic density}: We have annihilation modes induced by gauge and Yukawa interactions to explain the relic density of DM:
$\Omega h^2\approx 0.12$~\cite{Ade:2013zuv}, and their relevant Lagrangian in basis of mass eigenstates is found to be~\footnote{In general the second term below is also proportional to  $M'_{12}$ and $M'_{13}$, but these contributions are negligibly small when we take $m_{\psi_{2,3}}$ is larger than $M_X$ by a few factor at least.}  
\begin{align}
-{\cal L}&=
\mathcal{L}_{Z'} +i\frac{M'_{11}}{v'_{2}} \bar X P_R X z'_{\varphi_2}
\left(+ \sum_{\beta =2,3} i\frac{M'_{1\beta}}{v'_{2}} \bar X P_R \psi_\beta z'_{\varphi_2}\right) 
\nonumber \\
&+(Y_\nu)_{\alpha 1}\bar \nu_\alpha P_R X(\eta_R-i\eta_I)+\sqrt2 (Y_\nu)_{\alpha 1}\bar \ell_\alpha P_R X\eta^-+{\rm c.c.},
%%%
\end{align}
where $M'_{11}\equiv \sum_{i,j=1,2}V_{N_{1i}} y'_{N_{ij}} V^T_{N_{j1}} v'_{2}/\sqrt2$, and $f_{SM}$ is all the fermions of SM.
%%%
However since the typical Yukawa couplings in order to satisfy LFV constraints are ${\cal O}(0.01)$ as we see that in the previous subsection, one finds that any annihilation modes via Yukawa couplings cannot be dominant. Thus we just focus on the processes of $Z'$ and the GB final state.
%%%
Then the squared amplitudes for the processes
%%%
$X\bar X\to f\bar f,\nu_a\bar\nu_a, 2z'_{\varphi_2}$ are respectively given by
\begin{align}
|\bar {\cal M} (X\bar X\to f\bar f)|^2 & \approx 
\frac18 (s-4M_X^2) \sum_{f}
\left|\frac{g_{BL}^2Q^X_{BL}Q^f_{BL} }{s-m_{Z'}^2 + i m_{Z'} \Gamma_{Z'}}\right|^2 
\left[\cos2\theta(s-4m_f^2)+4m_f^2+3s\right], \\
%%%
|\bar {\cal M} (X\bar X\to \nu\bar\nu)|^2 & \approx 
\frac {3s}{16} (s-4M_X^2) %\sum_{f}
\left|\frac{g_{BL}^2Q^X_{BL} }{s-m_{Z'}^2 + i m_{Z'} \Gamma_{Z'}}\right|^2 (\cos2\theta+3), \\
|\bar {\cal M} (X\bar X\to 2{\rm GB})|^2 & \approx  -\frac{|M'_{11}|^4}{2 v'_{\varphi_2}}\times\nn\\&\hspace{-3cm}
\frac{2(2 M_X^4+ 2M_X^2 s-s^2)s^2 + s \cos^2\theta(s-4M_X^2)[s(s+4M_{X}^2)-4M_X^4+s\cos^2\theta(s-4M_X^2)]} 
{[s^2 - s\cos^2\theta (s-4M_X^2)]^2},
%%%
%|\bar {\cal M} (X\bar X\to 2{\rm GB})|^2 & \approx  -\sum_{\beta=1}^3 \frac{|M'_{1\beta}|^4}{2 v'_{\varphi_2}}\times\nn\\&\hspace{-5cm}
%\frac{2(2 M_X^4+ 2M_X^2 s-s^2)(2 M_{\psi_\beta}^2 -2M_X^2+s)^2 + s \cos^2\theta(s-4M_X^2)[s(4M_{\psi_\beta}^2+s)-4M_X^4+s\cos^2\theta(s-4M_X^2)]} {[(s-2M_X^2+ 2 M_{\psi_\beta}^2)^2 - s\cos^2\theta (s-4M_X^2)]^2},
\end{align}
where $s$ denote one of the Mandelstam variables, $\theta$ is one of the phase space angle which is integrated out from zero to $\pi$ as we will see below.
%%%
$\Gamma_{Z'}$ is the total decay width of $Z'$, where contributions from all SM fermions are included since we expect $m_{Z'}$ is rather heavy.   
The total decay width of $Z'$ is given by summing up the partial decay widths Eqs.~(\ref{eq:width1})-(\ref{eq:width2}) if kinematically allowed.
Note also that $Z'$ mass is given by $m_{Z'} = g_{BL} \sqrt{(v'_1)^2 + (8 v'_2)^2}$.
%%%
\if0
Similarly we obtain the squared amplitude for the process $X\bar X\to \ell_a\bar\ell_b$ as follows:
\begin{align}
|\bar {\cal M}(X\bar X\to\nu_a\bar \nu_b)|^2 & \approx 4\left|\frac{(Y_\nu)_{a1} (Y_\nu^\dag)_{1b}}{t-m_{\eta_0}^2}\right|^2(p_1\cdot k_1)(p_2\cdot k_2)
+4\left|\frac{(Y_\nu)_{a1} (Y_\nu^\dag)_{1b}}{u-m_{\eta_0}^2}\right|^2(p_1\cdot k_2)(p_2\cdot k_1)\nn\\
%
&
-\frac18 \left(\frac{(Y_\nu)_{a1} (Y_\nu^\dag)_{1b}}{t-m_{\eta_0}^2} \frac{(Y_\nu^\dag)_{1a} (Y_\nu)_{b1}}{u-m_{\eta_0}^2}
+
\frac{(Y_\nu^\dag)_{1a} (Y_\nu)_{b1}}{t-m_{\eta_0}^2} \frac{(Y_\nu)_{a1} (Y_\nu^\dag)_{1b}}{u-m_{\eta_0}^2}
\right) M_X^2 (k_1\cdot k_2)\nn\\
%
&+2\delta_{ab} \left|\frac{g_{BL}^2\left[-4+9 (V_N^*)_{13} (V_N^T)_{31}\right] }{s-m_{Z'}^2 + i m_{Z'} \Gamma_{Z'}}\right|^2 G(M_X, m_{Z'}, \{p_i,k_i\}) 
%
\end{align}
where we also assume massless final states for neutrinos. 
%and each of inner products such as $(p_1\cdot p_2)$ is found to be ref.~\cite{Cheung:2016ypw}.
In addition the squared amplitude for the process $X\bar X\to q_a\bar q_a$, which consists of $s$-channel, is given by
\begin{align}
&|\bar {\cal M}(X\bar X\to q_a\bar q_a)|^2\approx 
\frac23 \left|\frac{g_{BL}^2\left[-4+9 (V_N^*)_{13} (V_N^T)_{31}\right] }{s-m_{Z'}^2 + i m_{Z'} \Gamma_{Z'}}\right|^2 G(M_X, m_{Z'}, \{p_i,k_i\}),
\end{align}
\fi
%%%
Then the relic density of DM is given by~\cite{Edsjo:1997bg}
\begin{align}
&\Omega h^2
\approx 
\frac{1.07\times10^9}{\sqrt{g_*(x_f)}M_{Pl} J(x_f)[{\rm GeV}]},
\label{eq:relic-deff}
\end{align}
where $g^*(x_f\approx25)$ is the degrees of freedom for relativistic particles at temperature $T_f = M_X/x_f$, $M_{Pl}\approx 1.22\times 10^{19}$ GeV,
and $J(x_f) (\equiv \int_{x_f}^\infty dx \frac{\langle \sigma v_{\rm rel}\rangle}{x^2})$ is given by~\cite{Nishiwaki:2015iqa}
\begin{align}
J(x_f)&=\int_{x_f}^\infty dx\left[ \frac{\int_{4M_X^2}^\infty ds\sqrt{s-4 M_X^2}[ W^{f}(s)+W^{\nu}(s)+W^{\rm GB}(s)] K_1\left(\frac{\sqrt{s}}{M_X} x\right)}{16  M_X^5 x [K_2(x)]^2}\right],\\ 
%%%%%%%%%%%
W^f(s)&=
\frac{s-4M_X^2}{24\pi} \sum_{f} C_f
\left|\frac{g_{BL}^2Q^X_{BL}Q^f_{BL} }{s-m_{Z'}^2 + i m_{Z'} \Gamma_{Z'}}\right|^2 
\sqrt{1-\frac{4m_f^2}{s}} (2m_f^2+s),\\
%%%%%%%%%%%
W^\nu(s) &=
\frac{s(s-4M_X^2)}{16\pi} %\sum_{f}
\left|\frac{g_{BL}^2Q^X_{BL} }{s-m_{Z'}^2 + i m_{Z'} \Gamma_{Z'}}\right|^2 ,\\
%%%%%%%%%%%
W^{\rm GB}(s) &=
\frac{|M'_{11}|^4}{64\pi v'^4_{\varphi_2}} 
%\times\nn\\&\hspace{-3cm}
\left[
(3s^2-4M_X^4) \left( \frac{\pi}{2sM_X^2}\sqrt{\frac{M_X^4}{4sM_X^2-s^2}} - \frac{\tan^{-1}\left[\frac{s-2M_X^2}{\sqrt{s(4M_X^2-s)}}\right]}{s^{3/2}\sqrt{4M_X^2-s}} \right)
-4\right],
\label{eq:relic-deff}
\end{align}
where $C_f$ ($C_f=1$ for leptons and $C_f=3$ for quarks) is color factor, $W(s)$ is defined by 
$\frac{1}{16\pi} \sum_a \int_0^\pi \sin\theta |\bar {\cal M}|^2$, and we implicitly impose the kinematical constraint above.
 %%%%
In fig.~\ref{fig:relic}, we show $M_X$ and relic density of DM, fixing the following parameters:
\footnote{Notice here that the mode of GB does not depend on the mass of $\psi_1$ and $\psi_2$ so much.}
%\footnote{In principle, one has to derive this mixing and their masses by diagonalizing $M_N$ in the neutral fermions. But here we expect any values can be taken, since all the mass parameters except the DM mass and its mixing are free.}:
\begin{align}
 & g_{BL}=|V_{N_{13}}|=0.0075,\ m_{Z'}=500\ {\rm GeV},  v'_{2} =500 \ {\rm GeV},\nn\\
&  |M'_{11}|=150 \ {\rm GeV}, \ 700\lesssim M_{\psi_{2,3}},
%M_{\psi_{2,3}=700 \ {\rm GeV}, \ M_{\psi_3}=1000 \ {\rm GeV},
 \end{align}
 where $v'_1 \sim 66 \ {\rm TeV}$ is found which is consistent with the LEP bound $g_{BL}/m_{Z'}=1/\sqrt{v'^2_{1}+(8v'_{2})^2}\le 1/(7\ {\rm TeV})$~\cite{Schael:2013ita}.}
%Each of the red, blue and black line represents relic density depending on the mass of $Z'$ with $20$ GeV, $200$ GeV, and $600$ GeV.
%%%%%%%
We find that observed relic density can be obtained for $m_{Z'} \sim 2 M_X$ due to the resonance enhancement of the annihilation cross section.
%%%
In addition It also suggests that relic density can be also explained by $XX \to 2 GB$ mode where $M_X \lesssim 100$ GeV is required  for 150 GeV $\le |M'_{1\beta}|$;
for larger $|M'_{1\beta}|$, heavier DM mass region is also allowed.
%%%
Here let us explore this behavior of relic density considring the properties of annihilation modes.
Increasing the DM mass, the cross section of $Z'$ exchanging mode simply decreases up to the resonant point, and then it simply starts to increases as approaching $M_X \sim m_{Z'}/2$. Therefore its relic density behaves as the opposite manner to the cross section.   
On the other hand the cross section for GB final states simply decreases when we increase the DM mass fixing other parameters.
Then this behavior increases the relic density of DM.   
%%%
In total the relic density starts to increase up to $M_X\lesssim 230$ GeV (before the resonant point), since the GB mode  contribution is stronger than the $Z'$ exchanging mode.
But once it reaches at around the resonant point, the $Z'$ mode contribution becomes stronger due to the resonance.
After that, it simply increases due to both of property, when the DM mass increases.
%Thus one can also say that there are no left-side of points of solution ($M_X\sim 90$ GeV in our figure) when 150 GeV $\le |M'_{1\beta}|$, when all the parameters except $|M'_{1\beta}|$ are fixed.

%------------------------------------------------------------------------
\begin{figure}[t]
\centering
\includegraphics[width=10cm]{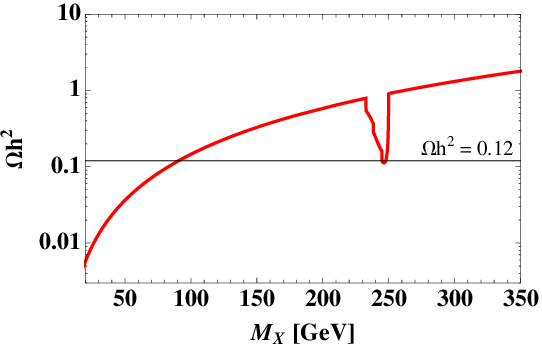}
\caption{The correlation between $M_{X}$ and $\Omega h^2$ for a benchmark point.  }
\label{fig:relic}
\end{figure}
%------------------------------------------------------------------------

As a comprehensive discussion, one might consider the case of coannihilation that can be possible in general.
The simplest case could be  that the masses of $\psi_{1,2,3}$ are almost degenerate.  In this case, the total cross section simply decrease.
Therefore, the allowed region at around the resonant point becomes to be narrower, and the lighter DM mass satisfying the relic density becomes to increase.
The other cases could cause among $Z'$ and or CP-even bosons, but this is beyond our scope because the behavior of relic density is very complicated.

\section{Conclusion}
We have proposed a model providing the neutrino mass and mixing at one loop-level with a nontrivial $U(1)_{B-L}$ gauge symmetry based on the model proposed by~\cite{Singirala:2017see},
in which the remnant $Z_2$ symmetry still be there even after the spontaneous symmetry breaking of $U(1)_{B-L}$,
and a fermionic DM candidate has been discussed instead of bosonic one.
Then we have given formulas for neutrino mass matrix, branching ratio of $\ell \to \ell' \gamma$ and relic density of DM.
Notice that, in our model, a physical GB appears, which is the consequence of two kinds of bosons $\varphi_1$ and $\varphi_2$ to break $U(1)_{B-L}$,
where we have selected $z'_{\varphi_2}$ as the physical GB by taking $v'_{2}<<v'_{1}$.
Then we have had a global analysis to satisfy the neutrino oscillation and LFVs, and found the typical order of Yukawa couplings are of the order 0.01. It suggests that Yukawa contribution to the DM relic density is negligibly small. Thus we have not considered this contribution in the DM analysis.
Instead, we have considered the GB contribution to the relic density of DM.
%However since it does not couple to any SM fields as well as DM directly, GB does not affect our phenomenology. 

In the DM analysis, we have shown the behavior of relic density in term os DM mass, preparing a benchmark point.
The first solution arises from the  contribution of GB mode, and the second one comes from $Z'$ mode as a resonance; $m_{Z'} \simeq 2 M_{X}$.

%\newpage
%%%%%%%%%%%%%%%%%%%%%%%%%%%%%%%%%%%
%\hspace{0.2cm} {\bf Acknowledgments}
%\section*{Acknowledgments}:
%\vspace{0.5cm}
\section*{Acknowledgments}
\vspace{0.5cm}
H. O. thanks Prof. Seungwon Baek for fruitful discussions, and
 is sincerely grateful for the KIAS member and all around.
%%%%%%%%%%%%%%%%%%%%%%%%%%%%%%%%%%%
%%%%%%%%%%%%%%%%%%%%%%%%%%%%%%%%%%%

\end{document}